# ON VARIATIONAL MICRO-MACRO MODELS AND THEIR APPLICATION TO POLYCRYSTALS

**Mayeul ARMINJON and Didier IMBAULT**

**Laboratoire "Sols, Solides, Structures" [associated with CNRS]**
**Institut de Mécanique de Grenoble, B. P. 53X, 38041 Grenoble cedex, France**

**Abstract**- Some variational micro-macro models are briefly reviewed: it is shown how, starting from the Taylor model and passing through the relaxed Taylor model, a consistent intermediate between Taylor's upper bound and the lower bound (Sachs or rather "static") model was obtained. This intermediate or "inhomogeneous" variational model (indeed, it generally predicts both strain and stress to be inhomogeneous) could offer a general alternative to self-consistent models. However, the implemented version was a rather empirical model (ARMINJON [1984]) with a less well-defined status. We present current progress in the implementation of the correct version.

## INTRODUCTION

As a model for micro-macro transition, the classical Taylor model is none other than the transposition to polycrystal plasticity of Voigt's uniform strain model, proposed for polycrystal elasticity: the new feature in Taylor's model was the principle of minimum plastic work. This "principle", and Hill's maximum work principle as well, were later recognized by BISHOP & HILL [1951] to be derivable properties of the single crystal constitutive law based on Schmid's law. They also proved that, under a no-correlation condition between the microscopic fields of stress and incremental strain, the maximum work principle, i.e. the normality flow rule, is transmitted from the crystal level to the polycrystal scale. These results were obtained without any recourse to the Taylor model, i.e. without assuming that the strain increment is uniform in the polycrystal. Yet in the same work, they proved that the latter assumption provides an upper bound to the polycrystal yield surface.

When extensive numerical calculations could be done, it was established that the main shortcomings of the Taylor model concern the prediction of *deformation textures*, which agree only qualitatively with the measured ones. This result is a particular case of a quite general feature in micro-macro transition- namely, that it is easier, hence more dangerous for the models, to consistently check the *localization* process (predict the micro-fields from the data of the overall stimulus) than to do so for the *homogenization* process (predict the overall response, by averaging the predicted field of microscopic response, found in the localization step). The explanation is in terms of a balance between input and output data. In order to run the model, one must know at least (i) the microscopic constitutive law, depending on the "state", i.e. on some list **X** of internal and geometrical variables, e.g. **X** = (**R**, ($\tau_s^c$)) with **R** the crystal orientation and $\tau_s^c$ ($1 \leq s \leq S$) the critical shear stresses, and (ii) the volume distribution of **X** (e.g. the ODF). In practice, often too little is known about this (e.g. little is known as regards the distribution of the critical shear stresses), so one has to make plausible assumptions which still leave some unknown parameters. But often also, few overall mechanical properties are known securely. The result is that the remaining unknown microstructural parameters can often be selected so as to get a reasonable agreement between measured and predicted overall properties (if one uses a reasonable model and if the inhomogeneity is moderated, as is the case for orientation-governed inhomogeneity in polycrystals; yet some models remain better than others, also for overall properties). In contrast, the knowledge of the microscopic fields gives access to a greater number of experimentally testable parameters. For example, an experimental ODF or texture function involves a good hundred parameters (texture coefficients). A polycrystal model allows to predict deformation textures, because it predicts the distribution of the velocity gradient among the different "grains" or crystal orientations.

## DEVELOPMENT OF SOME VARIATIONAL MODELS FROM THE TAYLOR MODEL

One successful attempt to improve the predictions of the Taylor model was the "relaxed Taylor model", which was built by several authors, beginning perhaps with RENOUARD & WINTENBERGER [1976] (although they did not actually propose a polycrystal model) and HONNEFF & MECKING [1978], and ending with VAN HOUTTE [1982]: this model indeed provided better predictions of deformation textures than did the Taylor model, especially for some cold-rolled f.c.c. materials (for a review, see VAN HOUTTE [1984]). It consists in leaving some shear strain-rate components "free" (grain-dependent), while the other components remain uniform as in the Taylor model; for every crystal orientation, the free strain-rate components are selected so as to minimize the work-rate. Thus, there are several versions of the model, according to the "free" components which are considered. However, in some cases, an improvement could be observed only if the predictions of several versions of the model were mixed. This appeared to be due to the simplistic schematization of the strain inhomogeneity: in reality, all strain-rate components are non-uniform (and in fact, within each grain as well) but, on the other hand, the inhomogeneity cannot reach arbitrarily high values- as is yet allowed to the free shear components in the relaxed Taylor model. Moreover, the "freedom" left to some components was justified by appealing to morphological considerations, which are suggestive but not perfectly clear and which, in any case, imply that the model should be applied only after large proportional strains have imposed a particular ("lath" or "pancake") form to the grains. For this reason, FORTUNIER & DRIVER [1987] suggested a model establishing a continuous transition from the Taylor model to the relaxed Taylor model. This model, however, still restricts the possible forms of strain-inhomogeneity and does not fully account for the fact that this is detected from the beginning, as it must be since a homogeneous strain is mechanically impossible in a heterogeneous material.

A different model was proposed by ARMINJON [1984], in which one feature of the relaxed Taylor model was conserved, i.e. the assumption that the strain-inhomogeneity obeys a principle of least energy expenditure, but the allowed strain-inhomogeneity was more general in form: it was only assumed that, for each grain $k$ ($k=1,...,n$), the norm of the strain-rate tensor $\mathbf{D}^k$ is the same ($|\mathbf{D}^k| = |\mathbf{D}|$) and $\mathbf{D}^k$ must lie within some ball centered at the overall strain-rate tensor $\mathbf{D}$, i.e. $|\mathbf{D}^k - \mathbf{D}| \leq \alpha$, where the number $\alpha$ is prescribed. ARMINJON [1987] determined the size of the ball so as to give the best prediction of the deformation texture for one deformation process (cold-rolling). The predicted texture was in very good agreement with the observed texture for cold-rolling; moreover, the predicted texture (with the same ball) agreed also very well with the experimental one for all other investigated deformations. These results were "explained" when this model was seen to be an approximation of another one, which is more rigorous- and also more general, since it may be applied to any material, provided the microscopic constitutive law derives from a potential. In this model (ARMINJON [1991a]), the independent energy minimization for every grain is replaced by a minimization of the average energy rate in the material under two constraints: (i) the average strain-rate is the overall strain-rate $\mathbf{D}$ ("consistency condition"), (ii) the average strain-rate inhomogeneity $h$ must not exceed a prescribed value $r$. It was proved mathematically that, for every $\mathbf{D}$, it indeed exists one value $r_0 = r_0(\mathbf{D})$ such that the so-obtained average energy rate $W_{r_0}(\mathbf{D})$ coincides with the actual overall energy rate $W(\mathbf{D})$. However, this does not justify the hypothese that the corresponding strain-rate *distribution* among the different constituents coincides with the actual one, $(\mathbf{D}^k)_{k=1,...,n}$. It was proved by ARMINJON et al. [1994] that the latter hypothese amounts to assume a precise principle of minimal inhomogeneity, according to which the average inhomogeneity $h$ of the actual strain-rate distribution is the least, among distributions satisfying the consistency condition (i) *and* leading to the actual overall energy rate $W(\mathbf{D})$.

This general model has been implemented for fiber-reinforced mortars, with few isotropic constituents, but has not been implemented for polycrystals, with many anisotropic constituents. The latter indeed did not seem urgent, in so far as the deformation textures were already accurately predicted with the 1984 approximate version. However, several observations indicate that the correct model should improve the predictions: e.g. the predictions of the Lankford coefficients $R$ are not better with the approximate version than with the Taylor model. The likely reason for this is that, in this version, each grain *separately* selects the strain-rate that minimizes the energy rate in the allowed ball, so that the consistency condition is not fulfilled. The $R$ coefficients and the whole flow rule depend sensitively on the shape of the yield surface, determinable if one knows the *ratios* of the critical shear stresses, usually assumed to be all equal to one for cubic materials. Hence they can be predicted from the texture function alone, thus providing an interesting test of polycrystal models, although they are overall properties.

### A NEW ALGORITHM FOR THE INHOMOGENEOUS VARIATIONAL MODEL

We envisage an aggregate with a finite number of constituents (zones with constant state, e.g. grains with given orientation), $k=1,...,n$, as will be the case in computational practice. However, *continuous* distributions of the states should rather be the generic case for a statistically homogeneous material (ARMINJON [1991a]). So the formulation should account for the fact that the finite number $n$ is the result of a discretization which may have to be refined so as to better approximate the continuous variation. Let us first state the "inhomogeneity constraint", $h \leq r$, in accordance with this requirement. The inhomogeneity $h$ of a distribution of the possible strain rates $\mathbf{D}^{*k}$ in the different constituents, is defined as the $p^{th}$-power average deviation of this distribution:

$$h = \left(\sum_{k=1}^{n} f_k \left|\mathbf{D}^{*k} - \mathbf{D}^*\right|^p \right)^{1/p} , \quad \mathbf{D}^* \equiv \sum_{k=1}^{n} f_k \mathbf{D}^{*k} , \qquad (1)$$

with $f_k$ the volume fraction of constituent $k$. The number $p \geq 1$ is not arbitrary, at least in the theoretical proofs (ARMINJON [1991a]): there it is determined with consideration of the behavior of the potential on straight lines $\mathbf{D} = \lambda \mathbf{D}_0$. In particular, for rate-independent plasticity, the potential is the work rate which is a homogeneous function of the strain-rate, and only $p = 1$ meets the conditions used in the proofs. In that case, one thus should not, strictly speaking, identify $h$ with the standard deviation (which corresponds to $p = 2$), as was occasionnally done for the purpose of simplicity in previous work. The consistency condition means that the average strain-rate, $\mathbf{D}^*$ in eqn (1), is assigned to be equal to the overall strain-rate $\mathbf{D}$. It is simply accounted for if the variable $\mathbf{Y}$ in the minimization problem is defined to be the sequence $\mathbf{Y} = (\mathbf{D}^{*k})_{k=1,\ldots,n-1}$, thus expressing the strain-rate in the last constituent as

$$\mathbf{D}^{*n} = \left(\mathbf{D} - \sum_{k=1}^{n-1} f_k \mathbf{D}^{*k}\right) / f_n . \qquad (2)$$

Since $f_1 + \ldots + f_n = 1$, we may then rewrite $h$ as

$$h(\mathbf{Y}) = \left(\sum_{k=1}^{n-1} f_k \left|\mathbf{D}^{*k} - \mathbf{D}\right|^p + f_n^{1-p} \left|\sum_{k=1}^{n-1} f_k (\mathbf{D}^{*k} - \mathbf{D})\right|^p \right)^{1/p} . \qquad (3)$$

Since the work rate functions $W_k$ of the constituents are convex, so is the function to be minimized, i.e. the average work rate:

$$F(\mathbf{Y}) = \sum_{k=1}^{n} f_k W_k(\mathbf{D}^{*k}) = \sum_{k=1}^{n-1} f_k W_k(\mathbf{D}^{*k}) + W_n(\mathbf{D} - \sum_{k=1}^{n-1} f_k \mathbf{D}^{*k}) . \qquad (4)$$

But the $W_k(\mathbf{D}^{*k})$ functions are not *strictly* convex, because each one is homogeneous, hence linear on straight lines $\mathbf{D}^{*k} = \lambda \mathbf{D}_0$ ($\lambda \geq 0$); so $F$ is not strictly convex. In the same way, $h(\mathbf{Y})^p$ is a convex function and it is strictly convex if $p>1$. The optimization problem is: find $\mathbf{Y}_0$ making $F(\mathbf{Y})$ a minimum, among the $\mathbf{Y}$'s satisfying the constraint $C(\mathbf{Y}) \equiv h(\mathbf{Y})^p - r^p \leq 0$. The saddle point theorem characterizes the solutions of this problem as follows:

$$\exists \lambda \geq 0 \text{ such that: } \nabla F(\mathbf{Y}_0) + \lambda \nabla C(\mathbf{Y}_0) = 0 \text{ and } \lambda C(\mathbf{Y}_0) = 0 . \qquad (5)$$

Hence, *unless $\lambda=0$* i.e. unless $r$ is so large ($r \geq r_S$) that the solution is that of the problem without constraint (which corresponds to the "static" model), then any solution satisfies the equality constraint $C(\mathbf{Y})=0$ i.e. belongs to the boundary of the convex set $\mathsf{C} = \{\mathbf{Y}; C(\mathbf{Y}) \leq 0\}$, furthermore the normal to $\mathsf{C}$ is also normal to the contour levels of $F$. As a consequence, if at least one of the two functions $F$ and $C$ is strictly convex, thus if $p>1$, then the solution $\mathbf{Y}_0$ is unique. Moreover, even if none is strictly convex, thus even if $p=1$, the non-uniqueness of the solution is likely to be an exceptional case. Unless $r$ is larger than the "static" value $r_S$, the numerical solution is searched as the solution of the problem with "penalized" equality constraint:

$$F_1(\mathbf{Y}) \equiv F(\mathbf{Y}) + F_\rho(\mathbf{Y}) = \text{Min}, \quad F_\rho(\mathbf{Y}) \equiv \rho C(\mathbf{Y})^2 , \qquad \rho \gg 1, \qquad (6)$$

and this solution is unique in any case. The gradient of $F_1$ involves the derivatives of $F$, which are easy to express in terms of those of the $W_k$ functions, i.e. the associated *stresses*:

$$\frac{\partial F}{\partial \mathbf{D}^{*k}} = f_k \left( \frac{\partial W_k}{\partial \mathbf{D}^{*k}} - \frac{\partial W_n}{\partial \mathbf{D}^{*n}} \right) = f_k \left( \sigma^k (\mathbf{D}^{*k}) - \sigma^n (\mathbf{D}^{*n}) \right) \quad (1 \leq k \leq n\text{-}1). \tag{7}$$

In passing, eqn (7) shows clearly that the minimum of the average work rate $F$ without inhomogeneity constraint ($\partial F / \partial \mathbf{D}^{*k} = 0$) indeed corresponds to the "static" model, i.e. to the case where the stress tensors in all constituents $k$ are (assumed to be) the same. It also implies that the second derivatives (making the Hessian matrix) are only numerically accessible, which led us to select the conjugate gradient method as the minimization algorithm for solving (6), instead of a quasi-Newton method like the BFGS method which was used by ARMINJON et al. [1994]. Moreover, these second derivatives are often irregular at points $\mathbf{Y}$ involving at least one $\mathbf{D}^{*k}$ which is normal to the corresponding yield surface ($\Sigma_k$) at a straight-line portion of ($\Sigma_k$), and are zero at vertices [that is, the stress tensor remains constant as long as the strain-rate lies within a cone of normals to ($\Sigma_k$)]. This also forbids the use of the Newton method even for unidirectional mimization in a known "descent" direction: therefore, a dichotomy is used for unidirectional minimization. Thus the usual Schmid law once more creates difficulties. For this optimization problem, the regular form proposed by ARMINJON [1991b] and by GAMBIN [1991] is probably preferable, and perhaps its physical content is not worse than that of Schmid's law.

## REFERENCES

1951 BISHOP, J.F.W. and HILL, R., "A Theory of the Plastic Distortion of a Polycrystalline Aggregate under Combined Stresses," Phil. Mag., **42**, 414.

1976 RENOUARD, M. and WINTENBERGER, M., "Déformation Homogène par Glissements de Dislocations de Monocristaux de Structure Cubique Faces Centrées sous l'Effet de Contraintes et de Déplacements Imposés," C. R. Acad. Sci. Paris **283 B**, 237.

1978 HONNEFF, H. and MECKING, H., "A Method for the Determination of the Active Slip Systems and Orientation Changes during Single Crystal Deformation," in GOTTSTEIN, G. and LÜCKE, K. (eds.), Textures of Materials (Proc. ICOTOM 6), Springer, Berlin- Heidelberg- New York, pp. 265-275.

1982 VAN HOUTTE, P., "On the Equivalence of the Relaxed Taylor Theory and the Bishop-Hill Theory for Partially Constrained Plastic Deformation of Crystals," Mater. Sci. Eng., **55**, 69.

1984 ARMINJON, M., "Explicit Relationships between Texture Coefficients and Three-Dimensional Yield Criteria of Metals," in BRAKMAN, C.M., et al. (eds.), Textures of Materials (Proc. ICOTOM 7), Netherlands Soc. Mater. Sci., Zwijndrecht, pp. 31-37.

1984 VAN HOUTTE, P., "Some Recent Developments in the Theories for Deformation Texture Prediction," in BRAKMAN, C.M., et al. (eds.), Textures of Materials (Proc. ICOTOM 7), Netherlands Soc. Mater. Sci., Zwijndrecht, pp. 7-23.

1987 ARMINJON, M., "Théorie d'une Classe de Modèles de Taylor 'Hétérogènes'. Application aux Textures de Déformation des Aciers," Acta Metall., **35**, 615.

1987 FORTUNIER, R. and DRIVER, J., "A Continuous Constraints Model for Large Strain Grain Deformations," Acta Metall., **35**, 509.

1991a ARMINJON, M., "Limit Distributions of the States and Homogenization in Random Media", Acta Mechanica, **88**, 27.

1991b ARMINJON, M., "A Regular Form of the Schmid Law. Application to the Ambiguity Problem," Textures and Microstructures, **14-18**, 1121.

1991 GAMBIN, W., "Crystal Plasticity Based on Yield Surfaces with Rounded-Off Corners," Z. angew. Math. Mech., **71**, T265.

1994 ARMINJON, M., CHAMBARD, T., and TURGEMAN, S., "Variational micro-macro transition, with application to reinforced mortars," Int. J. Solids Structures, **31**, 683.